\documentclass[twocolumn,preprintnumbers,superscriptaddress,landscape,nofootinbib]{revtex4}
\usepackage[utf8]{inputenc}
\usepackage{amsmath}
\usepackage{amsthm}
\usepackage{amssymb}
\usepackage{float}
\usepackage{braket}
\usepackage{graphicx}
\usepackage{url}
\usepackage{here}
\usepackage{natbib}
\usepackage{rotating}
\usepackage{comment}
\usepackage[colorlinks=true, pdfstartview=FitV, linkcolor=red, citecolor=blue, urlcolor=blue]{hyperref}
\usepackage{slashed}
\usepackage{multirow}
\usepackage[normalem]{ulem}
\graphicspath{{./figures/}}

\newcommand{\be}{\begin{equation}}      
\newcommand{\ee}{\end{equation}}      
\newcommand{\bea}{\begin{eqnarray}}      
\newcommand{\eea}{\end{eqnarray}}

\newcommand{\Tr}{\mathrm{Tr}}

\newcommand{\ctext}[1]{\raise0.2ex\hbox{\textcircled{\scriptsize{#1}}}}

\usepackage{tikz}
\usetikzlibrary{quantikz}

\theoremstyle{definition}

\theoremstyle{remark}

\usepackage[normalem]{ulem}

\begin{document}

\title{Investigating global and topological order of states by local measurement and classical communication: Study on SPT phase diagrams by quantum energy teleportation} 
\author{Kazuki Ikeda}
\email[]{kazuki7131@gmail.com}
\email[]{kazuki.ikeda@stonybrook.edu}
\affiliation{Co-design Center for Quantum Advantage, Department of Physics and Astronomy, Stony Brook University, Stony Brook, New York 11794-3800, USA}
\affiliation{Center for Nuclear Theory, Department of Physics and Astronomy, Stony Brook University, Stony Brook, New York 11794-3800, USA}

\bibliographystyle{unsrt}
\begin{abstract}
Distinguishing non-local orders, including global and topological orders of states through solely local measurements and classical communications (LOCC) is a highly non-trivial and challenging task since the topology of states is determined by the global characteristics of the many-body system, such as the system's symmetry and the topological space it is based on.
Here we report that we reproduced the phase diagram of Ising model and symmetry protected topological (SPT) phases using the quantum energy teleportation protocol, which foresees non-trivial energy transfer between remote observers using the entanglement nature of the ground state and LOCC. The model we use includes the Haldane model, the AKLT model and the Kitaev model. Therefore our method paves a new general experimental framework to determine and quantify phase transitions in various condensed matter physics and statistical mechanics. 
\end{abstract}

\maketitle

\section{How can we understand global or topological order of states by LOCC?}
To comprehend the characteristics of a quantum many-body system, it is crucial to grasp the global attributes of quantum states. Quantum states incorporate information about the symmetry, geometry, and topological structure of the system, which is important in determining the properties of matter~\cite{doi:10.1098/rspa.1984.0023,2017NatCo...8...50P,PhysRevX.8.031070,2018SciA....4.8685W}. However it is challenging to experimentally study an entire system of  many particles, and it remains a non-trivial question whether the properties of the entire system can be deduced from local observations only. In particular, since the topology and symmetry of a space cannot be determined by local properties alone, it is an open question how accurately the topological structure of a quantum state can be understood by local measurements of local observations. On the other hand, many experimental findings have shown that it is indeed possible to uncover the global properties of matter through local nature. 

One of the most successful such examples would be topological insulators~\cite{RevModPhys.88.035005,PhysRevX.7.041069}, exemplified by the quantum Hall effect, or more recently referred to as symmetry protected topological (SPT) states~\cite{PhysRevLett.61.2015,PhysRevLett.59.799,2001PhyU...44..131K,PhysRevB.81.064439,PhysRevB.85.075125,PhysRevB.83.075102,PhysRevB.72.045346,PhysRevLett.106.070501}. There remains much interest in the effect of local defects, impurities, and dislocations on the band structures~\cite{doi:10.1143/JPSJ.52.1740,Matsuki_2019,PhysRevB.104.035305,PhysRevLett.108.106403,PhysRevB.90.241403,PhysRevB.92.245139}. Furthermore, the duality between physical quantities in spaces of different dimensions, such as bulk-edge correspondence, is an interesting and profound problem associated with this puzzle~\cite{PhysRevLett.71.3697,PhysRevB.48.11851}. Such phenomena can be seen not only in condensed matter physics, but also in high-energy physics, as in the holography principle~\cite{1999IJTP...38.1113M,strominger2001ds,2022JHEP...05..129H,PhysRevLett.129.041601}. Therefore the inquiry of local vs. global aspects is a pivotal issue for a wide range of quantum physics. 

The main focus of this paper is to provide a universal approach to this local vs. global problem. To accomplish this, the most important concept is the entanglement of quantum states. In particular, the ground state entanglement is intrinsically important in understanding the global properties of the system using only local properties of states and physical quantities. This is because entanglement allows local information to be shared throughout the system, making it possible to access information in a subsystem from outside that subsystem. Therefore, it is reasonable to consider how to obtain information about the entire system from local operations, including measurements, using ground state entanglement as a strategy for the local vs. global problem.

As a protocol using only local operation and classical communication (LOCC), quantum teleportation is a very well-known and established protocol in quantum information theory and technology that allows the transfer of an unknown quantum state from one location to another~\cite{PhysRevLett.70.1895,furusawa1998unconditional,takeda2013deterministic,2015NaPho...9..641P}. By sharing an entangled state and exchanging classical information, a sender can teleport the quantum state to a receiver, who can then use local operations to recover the original state. This method enables the secure delivery of cryptographic keys to distant locations, thereby advancing the implementation of communication in quantum networks. It enables the secure quantum key distributions (QKD) \cite{2020arXiv200306557B}, which have led to the development of various applications, such as secure communication networks, secure authentication systems. These applications help to protect sensitive information and ensure the privacy and confidentiality of data transmission~\cite{farhi2012quantum,aaronson2009quantum,10.1007/978-3-030-01174-1_58,IKEDA2018199}.

While quantum teleportation can transmit quantum states, they are not physical quantities. For our ambitious goal of investigating the SPT phases by LOCC, we desire a way to access a physical quantity or non-local order at a remote location without direct observation. Is this really possible? Attempting this challenge in a classical system appears unwise because observing local quantities does not yield any information about the system as a whole.
However, the purpose of this paper is to present an affirmative and concrete solution to this question. For this, we use the protocol called quantum energy teleportation (QET) which was originally proposed as a means for remote observers to transfer energy via local computation and classical communication to their own subsystems~\cite{HOTTA20085671,2009JPSJ...78c4001H,2015JPhA...48q5302T,PhysRevA.84.032336}. Recently QET has been successfully implemented on real quantum hardware, showing excellent correspondence with the theoretical predictions~\cite{2023arXiv230102666I}. Moreover, the QET protocol is generalized so that it can be used for large-scale and long-distance energy teleportation by combining with quantum state teleportation~\cite{2023arXiv230111884I}.
In the previous paper by the author~\cite{2023arXiv230111712I}, the phase diagram of the QED$_2$ (two-dimensional quantum electrodynamics) was investigated by the QET protocol and it is reported that the teleported energy has a peak near the critical point of the phase transition. 

The purpose of this paper is to elucidate the link between SPT phases, entanglement entropy and QET. The phase diagrams computed reveal a striking correspondence for the critical points in both cases, implying that QET can be used a general mechanism to estimate any phase diagram and phase transition for SPT physics. A significant outcome of this will allow both theoretical and experimental work to be corroborated with the simulation, thus providing a simple method to diagnose phase transitions. Therefore, the QET protocol offers a wide range of benefit to explore quantum many-body systems by LOCC, including both high and low-energy physics.

\section{Quantum Energy Teleportation}
\begin{figure*}
    \centering
    \includegraphics[width=\linewidth]{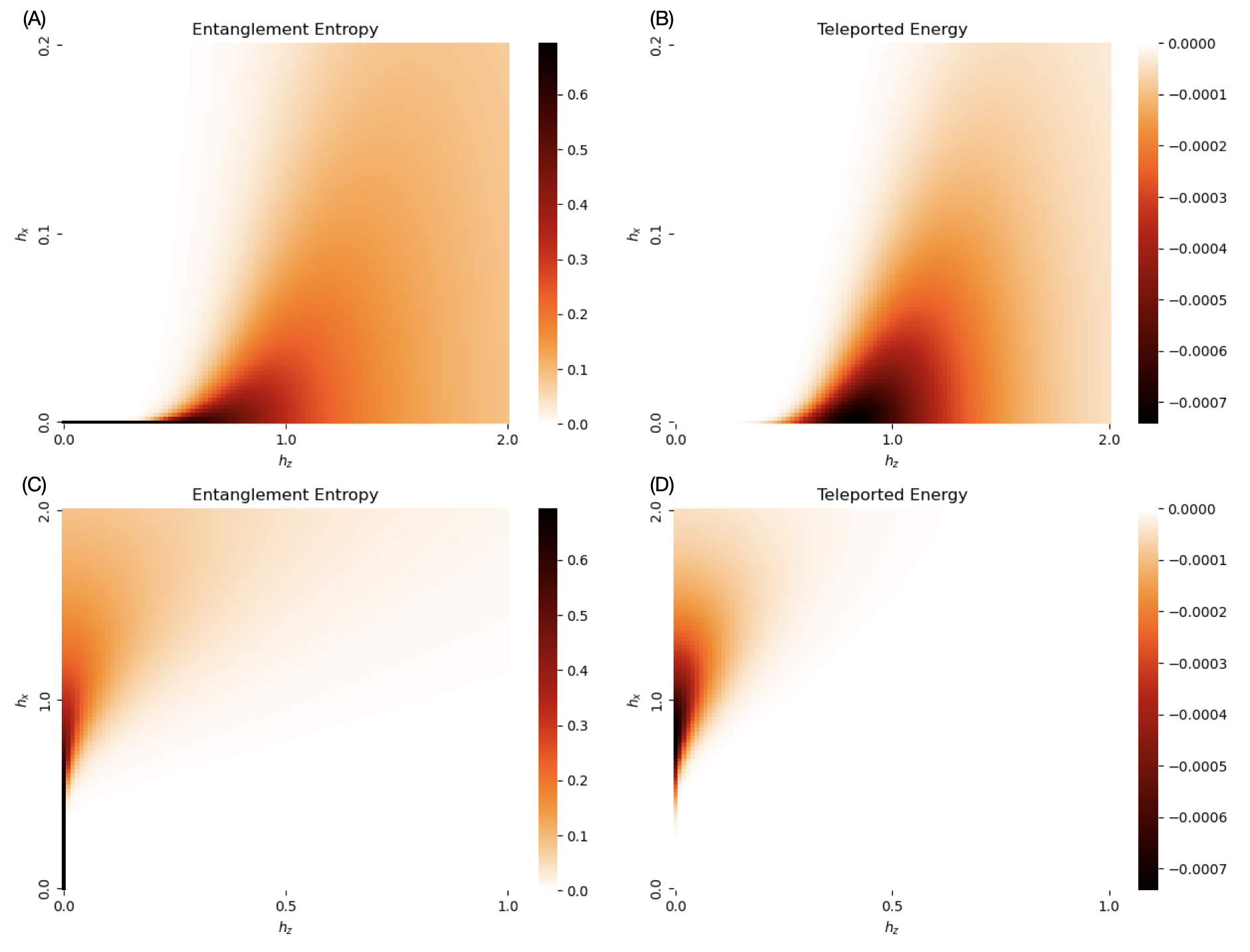}
    \caption{(A) Entanglement entropy of the Ising model with $\sigma_n=X_n$ and (B) the diagram of teleported energy.  (C)  Entanglement entropy of the Ising model with $\sigma_n=Y_n$ and (D) the diagram of teleported energy. Alice and Bob's operations $\sigma_{n_A}=Y_{n_A}$ and $\sigma_{n_B}=X_{n_B}$ were used for (B), and $\sigma_{n_A}=Y_{n_A}$ and $\sigma_{n_B}=Z$ were used for (C). $N=6,n_A=1,n_B=4$ were used.}
    \label{fig:ising}
\end{figure*}
\begin{figure*}
    \centering
    \includegraphics[width=\linewidth]{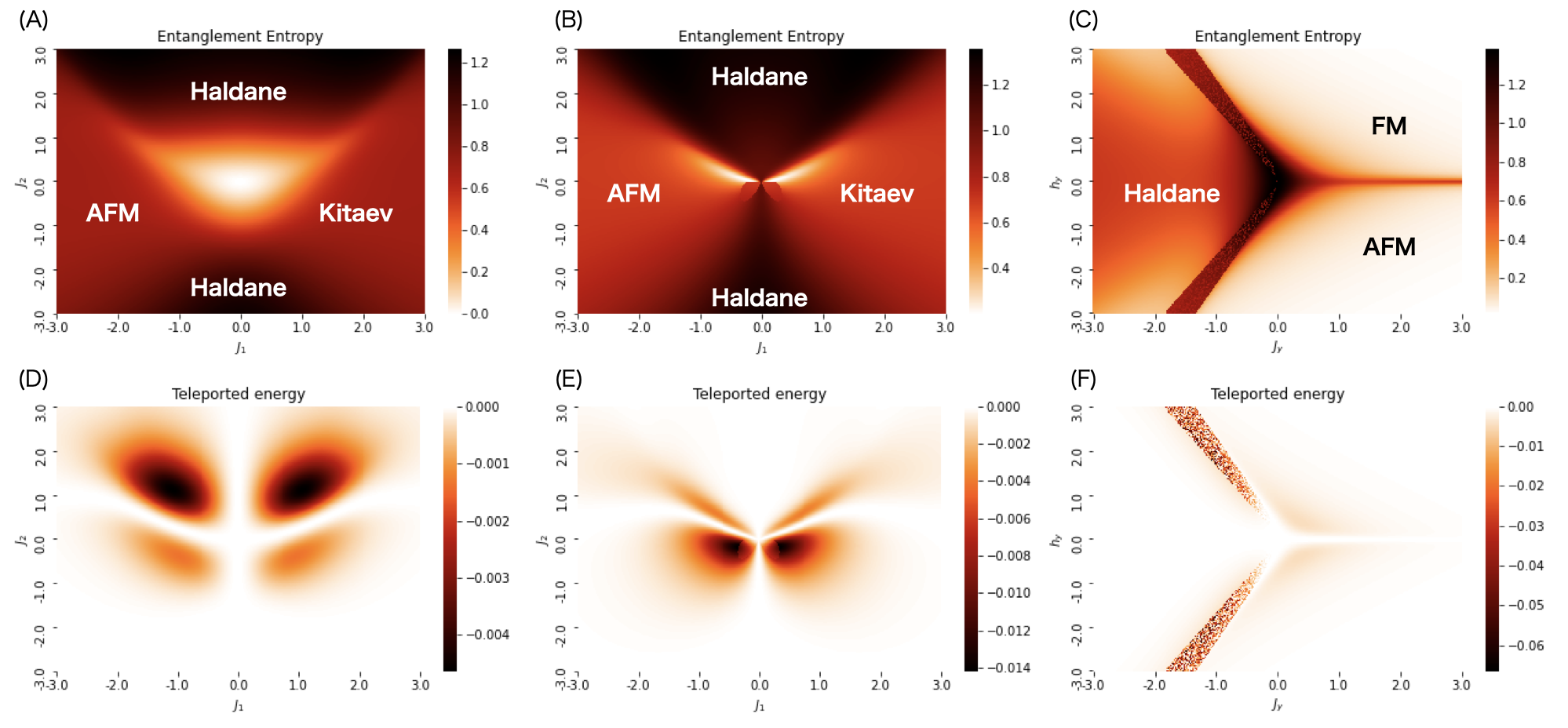}
    \caption{(A), (B), (C) Phase diagrams drawn by entanglement entropy. We generated (A) with the Hamiltonian \eqref{eq:Ham_spin}, (B) with the interaction term $\frac{1}{2}\sum_{n}Z_nZ_{n+1}$ and (C) with the Hamiltonian \eqref{eq:Ham_spin2}. Heatmaps of the teleproted energy are shown in (D), (E), (F), which were generated with the Hamiltonians that were used to generate (A), (B) and (C). For (D) and (E) we used $\sigma_{n_A}=X_{n_A}$ and $\sigma_{n_B}=Y_{n_B}$ for Alice and Bob's operations, and for (F) we used $\sigma_{n_A}=X_{n_A}$ and $\sigma_{n_B}=Z_{n_B}$. In all cases, we fixed the system size to $N=6$ and put Bob and Alice at $n_B=4$ and $n_A=1$ so that their coordinates have no overlaps.}
    \label{fig:HV}
\end{figure*}
Here we explain the general procedure of the QET protocol. Quantum circuit implementations of the protocol for various setups are given in~\cite{2023arXiv230102666I,2023arXiv230111884I,2023arXiv230111712I}. The same circuits provided in \cite{2023arXiv230111712I} work for this case as well, so we call this approach universal. Let us rewrite the Hamiltonian as $H=\sum_{n=0}^{N-1}H_n$,
where $H_n$ is the local Hamiltonian including the nearest neighbor interactions and obey the constraint
\begin{align}
\begin{aligned}
\label{eq:condition}
\bra{g}H\ket{g}=\bra{g}H_n\ket{g}=0,~\forall n\in \{0,1,\cdots,N-1\}
\end{aligned}    
\end{align}
where $\ket{g}$ is the ground state of the total Hamiltonian $H$. It is important to keep in mind that $\ket{g}$ is not always the ground state of local $H_n$. Non-trivial local operations, such as measuring the ground state, can result in excited states and increase the energy expectation value, which is supplied by the experimental setup. Additionally, $\ket{g}$ is usually an entangled state.

The steps of the QET protocol are as follows. Alice measures her Pauli operator $\sigma_{n_A}$ using $P_{n_A}(\mu)=\frac{1}{2}(1+\mu \sigma_{n_A})$ and gets either $\mu=-1$ or $\mu=+1$. Measuring the quantum state at subsystem $A$ disrupts the ground state entanglement and injects energy $E_A$ from the measurement device into the entire system. Initially, the injected energy $E_A$ is concentrated around subsystem $A$, but Alice cannot extract it from the system solely through her operations at $n_A$. This occurs due to the entanglement prior to the measurement causing information about $E_A$ to be stored in remote locations. To extract the energy from a location other than $n_A$, the quantum many-body nature of the system can be utilized through LOCC, as demonstrated later.

Via a classical channel, Alice sends her measurement result $\mu$ to Bob, who applies an operation $U_{n_B}(\mu)$ to his qubit and measures his local operators $X_{n_B},Y_{n_B},Z_{n_B}$ independently. The density matrix $\rho_\text{QET}$ after Bob operates $U_{n_B}(\mu)$ to $P_{n_A}(\mu)\ket{g}$ is 
where $\rho_\text{QET}$ is 
\begin{equation}
\label{eq:rho_QET}
    \rho_\text{QET}=\sum_{\mu\in\{-1,1\}}U_{n_B}(\mu)P_{n_A}(\mu)\ket{g}\bra{g}P_{n_A}(\mu)U^\dagger_{n_B}(\mu). 
\end{equation}
Using $\rho_\text{QET}$, the expected local energy at Bob's local system is evaluated as $\langle E_{n_B}\rangle=\Tr[\rho_\text{QET}H_{n_B}]$, which is negative in general. Due to the conservation of energy, $E_B=-\langle E_{n_B}\rangle (>0)$ is extracted from the system by the device that operates $U_{n_B}(\mu)$~\cite{PhysRevD.78.045006}. In this way, Alice and Bob can transfer the energy of the quantum system only by operations on their own lcoal systems.

It is beyond the scope of this paper to discuss how to actually extract this energy and the feasibility of doing so. Most importantly, the method for measuring $\langle E_{n_B}\rangle$ is already established and has been verified by real quantum hardware. In other words, the conclusion and results of this paper are not concerned with whether or not the positive net energy can be taken out of the system.

In what follows we give the details about the operations of Alice and Bob. We define $U_{n_B}(\mu)$ by 
\begin{equation}
    U_{n_B}(\mu)=\cos\theta I-i\mu\sin\theta\sigma_{n_B},
\end{equation}
where $\theta$ obeys 
\begin{align}
    \cos(2\theta)=\frac{\xi}{\sqrt{\xi^2+\eta^2}},~
    \sin(2\theta)=-\frac{\eta}{\sqrt{\xi^2+\eta^2}}
\end{align}
where
\begin{align}
    \xi=\bra{g}\sigma_{n_B}H\sigma_{n_B}\ket{g},~\eta=\bra{g}\sigma_{n_A}\dot{\sigma}_{n_B}\ket{g}
\end{align}
with $\dot{\sigma_{n_B}}=i[H,\sigma_{n_B}]$. The local Hamiltonian should be chosen so that $[H,\sigma_{n_B}]=[H_{n_B},\sigma_{n_B}]$. 
The average quantum state $\rho_\text{QET}$ is obtained after Bob operates $U_{n_{B}}(\mu)$ to $P_{n_A}(\mu)\ket{g}$. Then the average energy Bob measures is 
\begin{equation}
\label{eq:QET}
    \langle E_{n_B}\rangle=\Tr[\rho_\text{QET}H_{n_B}]=\frac{1}{2}\left[\xi-\sqrt{\xi^2+\eta^2}\right], 
\end{equation}
which is negative if $\eta\neq 0$. It is expected that if there is no energy dissipation, the positive energy of $-\langle E_{n_B\rangle}$ is transferred to Bob's device after the measurement due to energy conservation. 

Before we move on to the main discussion of the SPT phases, we present results of the Ising model with the transverse and the longitudinal fields
\begin{equation}
    H=-\sum_n[\sigma_{n}\sigma_{n+1}+h_xX_n+h_zZ_n+\epsilon_n],
\end{equation}
where $\sigma_n$ is either Pauli $X_n$ or $Z_n$ and $\epsilon_n$ must be chosen so that $H$ and $H_n$ respect the condition~\eqref{eq:condition}. We present the phase diagrams of this model in Fig.~\ref{fig:ising}. Figs.~\ref{fig:ising}A and B were drawn by $\sigma_n=X_n$, and Figs.~\ref{fig:ising}C and D were drawn by $\sigma_n=Y_n$. Figs. \ref{fig:ising}A and C correspond to the diagrams of the entangelement entropy and Figs. \ref{fig:ising}B and D are the heatmaps of the teleported energy. It is clear that those diagrams completely agree, although we use LOCC only to generate Figs.~\ref{fig:ising}B and \ref{fig:ising}D. These facts mean that long-range quantum correlations can be captured by LOCC alone, which is quite non-trivial and surprising. 

\section{Investigating SPT phase diagrams by energy teleportation}
There are a host of different one-dimensional non-trivial topological systems. These are highlighted by the SSH model, the AKLT model, and the Kitaev chain. Each of these models are ubiquitous in condensed matter physics. The SSH model consists of sites where the coupling is different on the odd sites compared to the even sites. This can be explicitly seen in the Hamiltonian 
\begin{align}
\begin{aligned}
\label{eq:Ham_spin}
\begin{split}
H_{\text{SSH}}&=2\sum_{n} (1-\lambda)c_{A,n}^{\dagger} c_{B,n} + \lambda c_{A,n+1}^{\dagger} c_{B,n}+ h.c.,
\end{split}
\end{aligned}
\end{align}
where $\lambda$ represents the coupling between the sites, and $c_i (c_{i}^{\dagger})$ denotes the annihilation (creation) operator. This is the simplest non-trivial topological model known. A similar but subtley different model is the Kitaev model
\begin{align}
\begin{aligned}
\label{eq:Ham_spin}
H_{\text{Kitaev}}&=\sum_{n} c_{n}^{\dagger} c_{n+1} + \lambda c_{n}^{\dagger} c_{n+1}^{\dagger} + h.c.
\end{aligned}
\end{align}
The major discovery linked to this model is related to Majorana physics being predicted to exist on the edge of this model. Additionally, we consider the AKLT model
\begin{align}
\begin{aligned}
\label{eq:Ham_spin}
H_{\text{AKLT}}&=\sum_{n} S_n S_{n+1} + \frac{1}{3} (S_n S_{n+1})^2,
\end{aligned}
\end{align}
where $S_i$ are spin-1 operators. Furthermore, the Haldane phase arises when there is a non-trivial topological order in the system, where the Haldane phase and AKLT model are intrinsically adiabatically connected.

It is worth emphasizing that these phases are only a small number of non-trivial phases. In order to determine how these different phases may arise due to varying interaction parameters, the phase diagram will be computed showing how the system can end up in one of these SPT phases. For the purpose of quantum simulation and measurement, we map the operators into the spin representation by the Jordan-Wigner transformation. The resulting spin Hamiltonians are given in eqs.~\eqref{eq:Ham_spin} and \eqref{eq:Ham_spin2}. See also~\cite{PhysRevB.96.165124} for those models. Interestingly, all of these phases can be witnessed by considering a single one-dimensional Hamiltonian, where the coupling between the sites is varied. We consider the Hamiltonian 
\begin{align}
\begin{aligned}
\label{eq:Ham_spin}
H&=\sum_{n}(Z_n-J_1X_nX_{n+1}-J_2X_{n-1}Z_{n}X_{n+1}+\epsilon_n),
\end{aligned}
\end{align}
where $J_1$ and $J_2$ are coupling parameters of the model. The Hamiltonian physically represents a combination of 1-,2-, and 3-spin interactions, where the spin interactions are dictated by the Pauli matrices. The benefit of considering this Hamiltonian is found in the rich structure of the phase diagram which yields a host of non-trivial phases. By varying these coupling parameters the phase diagram can be found, and from this it can be deduced how to interpret these different phases. Moreover, by having this phase diagram allows a direct comparison between this and the entanglement entropy, which is also plotted as a function of the coupling parameters. Utilizing this, can allow a direct understanding of how the ground state entanglement relates to the non-trivial topological phases that the system enters. The results for the teleported energy for eq. (\ref{eq:Ham_spin}) are shown in Fig.~\ref{fig:HV}E. It is worth emphasizing that eq. (\ref{eq:Ham_spin}) can be extended by including an interaction term given by $\frac{1}{2}\sum_{n}Z_nZ_{n+1}$. There is a first order phase transition between the AFM phase, and the Kitaev chain, whereas previously the transition was a second order phase transition. The difference is shown explicitly by comparing Fig. \ref{fig:HV}A with Fig. \ref{fig:HV}B and suggested in Fig. \ref{fig:HV}D and Fig. \ref{fig:HV}E. 

\begin{table*}
    \centering
    \begin{tabular}{|c|c|c|c|c|c|c|}\toprule
   $J_1$ &0.2 &0.4& 0.6& 0.8 &1.0\\\hline
$\langle X_{n_B-1}X_{n_B}\rangle$&$0.6030\pm0.0003$ &$0.8273\pm0.0002$&$0.8578\pm0.0002$ & $0.8811\pm0.0001$& $0.8996\pm0.0001$\\
    $\langle X_{n_B}X_{n_B+1}\rangle$&$0.1535\pm0.0003$ &$0.4572\pm0.0003$&$0.5772\pm0.0003$& $0.6631\pm0.0002$&$0.7276\pm0.0002$\\
    $\langle Z_{n_B-1}Z_{n_B}\rangle$ &$-0.0258\pm0.0003$&$-0.7111\pm0.0002$&$-0.5556\pm0.0003$& $-0.4364\pm0.0003$&$-0.3473\pm0.0003$\\
    $\langle Z_{n_B}Z_{n_B+1}\rangle$ &$0.6137\pm0.0002$&$-0.3115\pm0.0003$&$-0.2436\pm0.0003$& $-0.2000\pm0.0003$&$-0.1715\pm0.0003$\\
    $\langle Z_{n_B}\rangle$ &$-0.6162\pm0.0002$& $0.1834\pm0.0003$&$0.0523\pm0.0003$& $-0.0274\pm0.0003$ &$-0.0728\pm0.0003$\\\hline
    $\langle H_{n_B}\rangle$ &$-0.0021\pm0.0006$&$-0.0035\pm0.0006$& $-0.0031\pm0.0006$& $-0.0021\pm0.0006$ &$-0.0001\pm0.0006$\\
    $\langle H_{n_B}\rangle_\text{exact}$ &-0.0022&-0.0034& -0.0026& -0.0015&-0.0008\\\hline
    $\epsilon_{n_B}$ &0.4714&0.8382&1.2052& $1.5788$ &1.9593\\
    $\theta$ &-0.0486&0.0455& 0.0329& $0.0221$ &0.0145\\\hline
\end{tabular}
    \caption{Data from the quantum circuits. $J_2$ was put to 0 and each error corresponds to statistical error of $10^7$ shots. $\theta,\epsilon_n, \langle H_{n_B}\rangle_\text{exact}$ were computed by the exact method.}
    \label{tab:simulator}
\end{table*}
In order to further explore different SPT phases consider the Hamiltonian
\begin{align}
\begin{aligned}
\label{eq:Ham_spin2}
H&=\sum_{n}(h_yY_n-J_yY_nY_{n+1}-X_{n-1}Z_{n}X_{n+1}+\epsilon_n),
\end{aligned}
\end{align}
where $h_y$ and $J_y$ are coupling parameters for the system. Note that when $h_y=0$, this corresponds the free-fermion model. This model is similar as it includes a host of spin interactions, however now the focus is on the interactions due to $Y_n$ compared to $X_n$. The previous analysis can be applied to eq. (\ref{eq:Ham_spin2}) for this model in order to produce Fig.~\ref{fig:HV}C and Fig.~\ref{fig:HV}F. It is clear the underlying physics is fundamentally different, however the clear transitions between the phases is still visible. Therefore, it is apparent that utilizing QET allows an operational mechanism in order to approximate transitions in the SPT phases.

Now let us discuss the phase diagrams of the SPT states. In order to compare the results derived using QET, the ground state must be chosen in terms of the ground state such that the density matrix is defined by $(\rho=\ket{g}\bra{g})$. For each different Hamiltonian, there is a different ground state for the system. This can be seen in Figs.~\ref{fig:HV}A and B which show entanglement entropy $S(\rho)=-\Tr_L(\rho_R\log\rho_R)$ between the left (L) and right (R) half subsystems for the Hamiltonian~\eqref{eq:Ham_spin} and the that with the interaction term~$\frac{1}{2}\sum_{n}Z_nZ_{n+1}$. In addition, Fig.~\ref{fig:HV}C is the diagram generated with the Hamiltonian~\eqref{eq:Ham_spin2}. The figures exhibit sharp peaks at the critical points of phase transitions. Therefore, the link between ground state entanglement and phase transitions can be utilized in order to allow a comparison between the entanglement entropy and the QET protocol.

We performed a quantum simulation of QET in various SPT phases. Figs.~\ref{fig:HV}D, E and F show the teleported energy $\Tr[\rho_\text{QET}H_{n_B}]$ to Bob's local system. By studying Fig.~\ref{fig:HV}, it is clear that there is a direct correspondence between the entanglement entropy and the QET protocol. Furthermore, it is significant that the teleported energy is enhanced along the critical points of the phase transition. This suggests some underlying relation between phase transitions and the energy teleportation. Bob's local energy can be calculated by the explicit form of his local Hamiltonian. From those figures, it is clear that we have successfully reproduced the phase diagram of SPT phases using a quantum energy teleportation protocol and showed that it is possible to detect the topological properties of the entire system through LOCC. We demonstrate this through quantum simulation using the Haldane, AKLT, and Kitaev models. 

As a good consistency check with quantum circuits, we execute the program was executed by IBM qasm\_simulator. The quantum circuits are given in the previous work by the author~\cite{2023arXiv230111712I}. We focus on the Hamiltonian~\eqref{eq:Ham_spin2} with the interaction term $\sum_n\frac{1}{2}Z_nZ_{n+1}$, and examine the Kitaev phase, which is given in Fig.~\ref{fig:HV}B. The results are given in Table 1 and compared with the exact value. Bob's expectation value can be evaluated by $\langle H_{nB}\rangle=\langle Z_{nB}\rangle+\frac{1}{2}(\langle Z_{nB-1}Z_{nB}\rangle+\langle Z_{nB}Z_{nB+1}\rangle)-J_1(\langle X_{nB-1}X_{nB}\rangle+\langle X_{nB}X_{nB+1}\rangle)$. In all respects, agreement with the values using the universal formula~\eqref{eq:QET} was confirmed within the statistical error.

It is extremely non-trivial to determine the topological structure of a quantum state through local measurements of local observations since the topology and symmetry of a space cannot be fully determined by local properties alone. Nevertheless, we found that the entanglement of the ground state plays a crucial role in comprehending the global characteristics of a system through local states and physical quantities. This is due to the fact that entanglement enables the sharing of local information throughout the system, making it feasible to obtain information from a subsystem outside of it. 

\if{

A general definition for the change in entropy before and after the measurement by Alice is given by
\begin{align}
    \Delta S_{AB}=S_{AB}-\sum_{\mu}p_\mu S_{AB}(\mu)
\end{align}
where $p_\mu$ is the probability distribution of $\mu$,  $S_{AB}(\mu)$ is the entanglement entropy after the measurement, $\xi=\arctan\left(\frac{k}{h}\right)$. It is natural to expect the entanglement properties of the system to be modified after measurement due to the destructive effect that measurement often has on entanglement. After Alice's post-measurement, the state is mapped to 
\begin{equation}
    \ket{A(\mu)}=\frac{1}{\sqrt{p_\mu}}P_{n_A}(\mu)\ket{g}. 
\end{equation}
Then $S_{AB}(\mu)$ is calculated with the density matrix $\ket{A(\mu)}\bra{A(\mu)}$.
The analysis derived here corroborates the analysis conducted in ~\cite{2011arXiv1101.3954H,2009JPSJ...78c4001H}, where $\Delta S_{AB}$ is found to be the upper bound of a function $f(\xi,\eta)$ in such a way that
\begin{equation}
    \Delta S_{AB}\ge f(\xi,\eta)E_B.
\end{equation}
This gives a direct relation between the entanglement entropy and the teleported energy. Whilst, this inequality doesn't give an exact relation, it still confirms the results in Fig.1 that there is an underlying physical mechanism between ground state entanglement and QET.
It is important to emphasize that Bob's conditional operations reduce the entropy of his local subsystem. If Bob does nothing after Alice's measurement, then Figs~\ref{fig:HV}~(D) and (E) reveal how QET can yield the entanglement entropy after allowing the evolution of the system, such that entanglement arises.

An analogous expression to the one derived above reveals that there must exist some maximum energy that can be teleported which will be related to the change in entanglement entropy. This is given by
\begin{equation}
    \max_{U_1(\mu)}E_B\ge h(\xi,\eta)\Delta S_{AB},
\end{equation}
where $h(\xi,\eta)$ is a certain function. Given it is apparent that there is a clear connection between QET and entanglement entropy, this implies that there should be a universal relation between these two quantities, which allows a direct mapping from one to the other. The underlying similarity between these two quantities is the entanglement of the initial state, therefore this type of analysis may also provide new theories which have benefit in quantum information theory.

\url{https://online.kitp.ucsb.edu/online/compqcm10/oshikawa/pdf/Oshikawa_CompQCM.pdf}
}\fi

\section{Conclusion and Discussion}
The approach we propose offers a comprehensive experimental framework for identifying and measuring phase transitions using only LOCC in various fields of condensed matter physics and statistical mechanics. It would be worthwhile for future research to verify whether this method works well for SPT states in more than one dimension. In the previous paper~\cite{2023arXiv230111712I} and in this paper, we suggested that the QET protocol is effective in detecting phase transitions using only LOCC for a wide range of locally interacting models. In particular, as mentioned in the introduction, it would be very useful to consider the band theory of two-dimensional Riemann surfaces to achieve the goal of revealing the geometry or topology of the system through LOCC. The curvature of Riemann surfaces is divided by the number of species and is known to be positive for $g=0$, zero for $g=1$, and negative for $g\ge 2$. Riemann surfaces with $g\ge2$ are obtained by introducing appropriate translation symmetries, in the Poincar\'{e} disk and by properly gluing the edges of hyperbolic lattices together. Testing our method using hyperbolic band theory~\cite{doi:10.1126/sciadv.abe9170,doi:10.1073/pnas.2116869119,2021arXiv210710586I,Ikeda_2021,KIENZLE2022108664} will be a new and interesting problem that spans algebraic geometry, quantum information, and condensed matter physics. In particular, it is very interesting if one can distinguish between the $g=1$ and $g=2$ topologies, which is equivalent to distinguishing between zero and negative curvature of geometry using only LOCCs. An environment is already in place to conduct experiments using a real quantum device designed for hyperbolic lattices~\cite{kollar2019hyperbolic}.

\section*{Acknowledgement}
I thank Adrien Florio, Fangcheng He, Sebastian Grieninger, Dmitri Kharzeev, Qiang Li, Steven Rayan, Shuzhe Shi, Rajeev Singh, Hiroki Sukeno, Sergey Syritsyn, Derek Teaney, Jacobus Verbaarschot, Tzu-Chieh Wei, and Ismail Zahed for fruitful communication and collaboration. I also thank Adam Lowe for his useful comments on the draft. I thank IBM for the use of the simulator. The work was supported by the U.S. Department of Energy, Office of Science, National Quantum Information Science Research Centers, Co-design Center for Quantum Advantage (C2QA) under Contract No.DESC0012704.

\section*{Author contributions}
All work was performed by the author.

\section*{Competing interests}
The author declares that there are no competing financial interests. 

\section*{References}
\bibliographystyle{apsrev4-1.bst}
\bibliography{ref}
\end{document}